# ALIGNMENT OF SPEECH TO HIGHLY IMPERFECT TEXT TRANSCRIPTIONS

*Alexander Haubold and John R. Kender*

Department of Computer Science, Columbia University

## ABSTRACT

We introduce a novel and inexpensive approach for the temporal alignment of speech to highly imperfect transcripts from automatic speech recognition (ASR). Transcripts are generated for extended lecture and presentation videos, which in some cases feature more than 30 speakers with different accents, resulting in highly varying transcription qualities. In our approach we detect a subset of phonemes in the speech track, and align them to the sequence of phonemes extracted from the transcript. We report on the results for 4 speech-transcript sets ranging from 22 to 108 minutes. The alignment performance is promising, showing a correct matching of phonemes within 10, 20, 30 second error margins for more than 60%, 75%, 90% of text, respectively, on average.

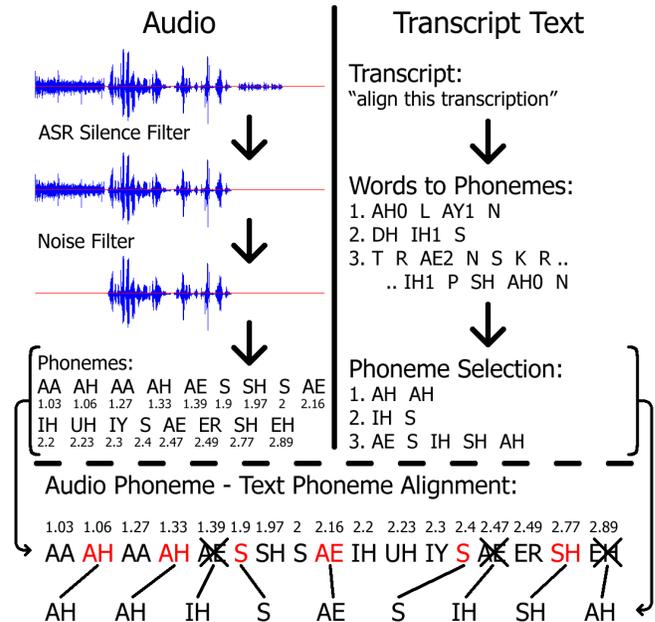

Figure 1. Overview of speech-text alignment for a sample phrase (without errors). Audio is filtered and selected phonemes are extracted using formant estimation or dominant frequency readings from a spectrogram. Temporally unaligned text (perfect or imperfect) is converted to phonemes. Alignment is performed using dynamic programming edit-distance transformation.

## 1. INTRODUCTION

With the growing use of multimedia as a medium for communication, delivery, and archiving, low cost systems for their automatic indexing become more widespread. Where previously costly camera setups and diligent post-processing were utilized for selected events such as large lectures, inexpensive cameras with amateur operators are now used to film even the smallest events, for example, student presentations in a classroom. The utility of such footage is inarguably justified, in particular for the students and instructors reviewing student performance [1]. However, the additional effort to produce accurate indices is prohibitively costly. Relying on automatic tools to generate approximations would be sufficient and helpful enough to manage large growing archives of video (100's of hours).

Speech transcriptions for semi-professional video productions, such as lectures, have traditionally been a manual task performed by commercial transcription services. Depending on the invested effort, such transcripts are either perfect or approximate. With the introduction of ASR systems, similar approximate transcriptions can be produced, especially when speaker and language models are appropriately created. In a classroom environment with more than one hundred students exhibiting a wide variety of accents and speech qualities, training becomes an infeasible task. Relying on commercial software packages such as IBM ViaVoice is an easy and inexpensive alternative. However, their untrained use in a sub-optimal recording environment comes with a severe trade-off in recognition quality and lack of text-to-speech alignment. Word error rates (WER) are commonly as high as 75% with word recognition accuracy around 40%. Despite the low accuracy, we have previously shown that for keyword indexing and searching, this quality is sufficient [1].

A further shortcoming of IBM ViaVoice and other commercially available recognition systems is their lack of time stamps for the recognized text. The missing temporal alignment for finding locations of interest specified by keywords hinders the user from making full use of the search capability. In general, a cheap ASR transcript combined with a cheap post-processing method is likely a cost effective approach. We address this problem of aligning highly imperfect transcripts to their original speech signal by relying on a class of phonemes which are typically recognized with the most accuracy. We apply our approach to lecture videos and student presentation videos for indexing and archiving.

| | |
|---|---|
| 1950: **for a way** in which people can tell a video is different from another video based on **the fact that** they are simply different faces you have to be aware of these | |
| 1960: of things you wanna retrieve *according* to some sort of *deep* **structure** formalism so *for* **example range of** color or | |
| 1950: **for a way** content **the fact that** can be captured a | |
| 1960: with and one of the true *accounting* deferrals are *indeed* **structure** plan live on *or* **example range of** colors but | |

Table 1. Top: manual (perfect), Bottom: automatically generated (highly imperfect) transcriptions. Timestamps precede the phrases. Highlighted are **correctly identified words**, correctly identified stems, and *incorrect, but similarly sounding words*.

## 2. RELATED WORK

While automatic alignment of text to speech has received some attention for different research goals, the topic of aligning highly imperfect transcript text to speech has not been covered in much detail. Common to related work is the use of moderately imperfect transcripts, in which WER is below 25%. The authors of [2] apply time-aligned ASR transcriptions to manually generated transcripts by selecting words as anchor points. While this approach is applicable to low-error text in which most words are correctly matched, we are not able to do so for highly imperfect transcripts. In [3] the author discusses alignment of ASR speech transcriptions to manual transcriptions for improving the combined WER. This approach makes use of ASR text which is generated with time codes. WinPitchPro, a tool presented in [4], features manual alignment of text to speech. Speech is played back at a reduced rate while the user clicks on matching words from the existing transcript.

Our approach to text to speech alignment shares several methods with [2]. The novelty of our approach lies in producing an alignment for a target corpus of highly imperfect transcripts. We perform the alignment on selected phonemes instead of on words. The resulting alignment allows us to generate a UI to locate specific parts of a lecture video or student presentations. (We have experimented with other approaches, including text-to-speech alignment based on a selected set of commonly occurring anchor words. However, speech qualities varied greatly between speech training samples, introducing too large of an error on the speech test set.)

## 3. APPROACH

Our approach addresses several shortcomings of the speech and text in our data set. Speech data is taken from audio sources with varying qualities in compression and recording environment. Audio from lectures is obtained from highly compressed videos, in which the instructors' audio quality is constant throughout. Audio from student presentations is taken from much lower compressed videos (DV tapes), where recording qualities are not constant. Besides highly varying speech and language qualities, different speakers (students) have different presentation characteristics.

Transcripts are obtained through the IBM ViaVoice ASR engine without specially designed language or speaker models. The resulting text exhibits a typical WER of over 70% (computed from edit distance) with the number of correctly identified words below 40%. Depending on an individual speaker's characteristic, a transcription can be reasonably accurate or filled entirely with noise. In some cases where the recording quality or the speed of speakers' presentation changes dramatically, we observe cases where the ASR engine skips several words at a time. An example of a typical transcription is presented in Table 1.

Due to these varying characteristics, an approximate temporal alignment between text and speech using easily obtained parameters, e.g. signal strength, is not favorable. We have observed through experimentation, that while transcriptions contain significant word errors, vowels and fricatives tend to be recognized with high accuracy and can still be found in the incorrectly recognized words. Our approach takes advantage of this observation, and performs text to speech alignment based on the selective subset of monophthongs (single vowels) and fricatives.

We first filter audio data for regions that do not correspond to speech, or are unlikely to be transcribed by the ASR engine due to poor audio quality. Phoneme detection is performed on the filtered signal, resulting in a temporally accurate list of speech phonemes. We tokenize the unaligned transcript into phonemes using a dictionary. Speech phonemes are then aligned to text phonemes using dynamic programming, resulting in a temporal alignment which can be mapped to the original transcript.

### 3.1. Phoneme Extraction from Audio

The goal of phoneme extraction is to assemble a list of likely occurring phonemes in the audio signal (Table 2), which can later be aligned to text phonemes.

Monophthong detection takes advantage of the characteristic stable voicing for vowels, which can be estimated by formant frequencies. The Matlab-based toolkit introduced in [5] models the vocal tract using an autoregressive model of the speech signal. Peaks in the frequency response correspond to the resonant frequencies of the vocal tract, or formants. Using a table of expected frequency values for formants F1, F2, and F3, the closest-matching phoneme is determined by the Euclidean distance of a weighted difference between model and computed values. We ignore any detected phonemes beyond an experimentally determined threshold.

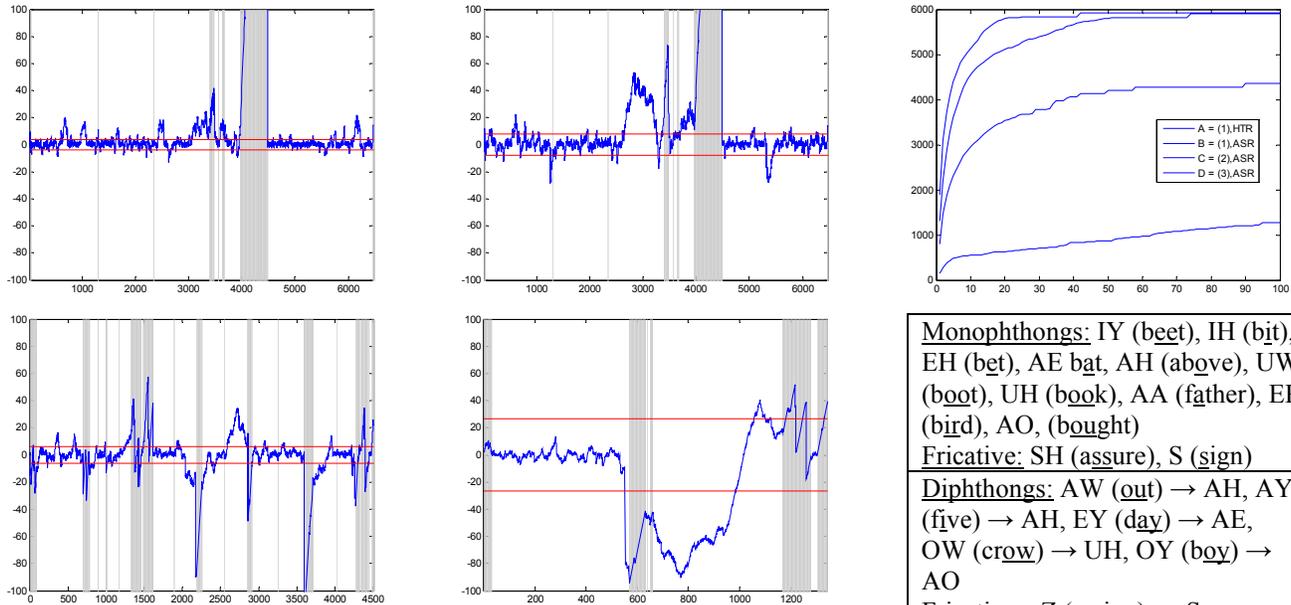

Figures 2a-e. (a,b,d,e) Alignment errors for data sets A, B, C, D. The x-axis represents length of the audio (sec), the y-axis error of phonemes' computed alignment. Horizontal lines mark the average error. Spikes generally co-occur with regions of silence or noise (shaded vertical bars). (c) Correct speech-text alignment for error margins 1-100 seconds (x-axis), and cumulative time of audio file. More than 75% of correct alignment occurs within a 20 second error margin.

Table 2. (top) Phonemes detected in speech and text; (bottom) Rules of substitution for phonemes not specifically detected in speech, but exhibiting similarities.

Detection of fricatives is highly dependent on the distribution of energy in frequency bands illustrated by spectrograms. We use the expected distributions of energy among frequency bands to detect the fricatives SH, and S. For a given window of speech signal, we select the maximum value of normalized cumulative energy in the expected frequency bands: SH/CH = [2500-3000 Hz], S/Z = [3000-4000 Hz], all others = [300-2500 Hz].

Phoneme detection is performed on small windows of the audio signal of 1/30th of a second in length. This window is intentionally smaller than the average duration of a phoneme in particular vowels. In a final step, neighboring phonemes of the same type are merged.

### 3.2. Phoneme Extraction from Text

In this step we generate a collection of phonemes from the unaligned transcript. While pronunciation of words depends largely on dialect, it is infeasible to tune the phonetic dictionary as a dialect model, in particular when an audio track features 30 or more speakers in short intervals. We assume pronunciation for American English, and make use of the CMU Pronouncing Dictionary with over 125,000 words and their phonetic transcriptions [6].

Text is segmented into words, which are then represented by their phonemes. Words not found in the dictionary are shortened to find the closest stem. Numerical values are converted to their verbal counterparts, treating the pronunciation of digits differently from that of compound numbers. We apply a set of rules for several phonemes which are not detected during speech phoneme extraction, but which share phonetic features with the identified monophthongs (Table 2). The final representation of a word includes only phonemes detected by speech.

### 3.3. Alignment

We perform alignment globally between the sets of phonemes from speech and text using the edit-distance dynamic programming algorithm, similar to the alignment task between two DNA sequences. Our edit-distance implementation aligns text phonemes to a larger set of speech phonemes by incurring copies, deletions, insertions, and replacements, but not transpositions (Table 4). Because of the significantly redundant set of speech phonemes, the cost of incurring copies and deletions (-1) are the same, while replacements and insertions are assigned a cost of +1.

A typical set of speech phonemes for 60 minutes of audio contains up to 45,000, while the equivalent set of text phonemes contains up to 15,000 elements (Table 4). Once completed, time codes from individual phoneme matches are assigned to their original words, thus producing the temporal text to speech alignment for the user interface.

|   | Audio Source | Length | Quality | Features |
|---|---|---|---|---|
| 1 | Lecture | 1:48:21 | 16 kHz | Single speaker, one long break (504 sec) |
| 2 | Student Team Presentation | 1:15:12 | 48 kHz | 31 speakers, 6 Q&A sessions of varying durations (30 – 300 sec) |
| 3 | Student Presentation | 0:22:32 | 48 kHz | 10 speakers, 2 Q&A session of durations 60 sec and 185 sec |

Table 3. Source of audio are lecture and student team presentation videos. The lecture video features one speaker (male) who presents continuously with constant audio quality. Student team presentations feature many speakers (5-6 male and female students per team) and audience members during Q&A sessions at highly varying audio and speech qualities.

|   | Video and Transcript | Avg. Matching Error (sec) | # Speech phonemes | # Text phonemes | Copies / Deletions / Insertions / Replacements in set of Speech phonemes |
|---|---|---|---|---|---|
| A | (1), HTR | 3.9 | 64265 | 27645 | 16695 / 39189 / 2569 / 8381 |
| B | (1), ASR | 7.7 | 64265 | 21608 | 14121 / 43997 / 1340 / 6147 |
| C | (2), ASR | 6.43 | 54537 | 16248 | 11459 / 38947 / 658 / 4131 |
| D | (3), ASR | 26.73 | 12596 | 4520 | 2960 / 8395 / 319 / 1241 |

Table 4. Speech-text alignment stats and results. (ASR = automatically generated, HTR = human generated transcripts)

## 4. EXPERIMENTAL RESULTS

We have conducted experiments with 3 speech files and 4 transcriptions, three of them highly imperfect and automatically generated, and one perfect and manually generated. Speech files were taken from lectures and student presentations (Table 3). We note that alignment is accurate within a reasonable error margin, and is sufficient to search a video stream. On average, more than 60%, 75%, 90% of all words are aligned correctly within a 10, 20, 30 second error margin, respectively. Figures 2a-e illustrate the phoneme alignment error for the data sets. The significant jump in Figures 2a and 2b are due to a silence break of more than 8 minutes in the extended lecture of 108 minutes. Similar spikes can be found in Figure 2c, where 5 Q&A periods between presentations cause phonemes close to the silenced break to be misaligned. The noticeably large error in Figure 2d is due to a number of factors related to speech quality, including increased speed of the student's speech and volume, both leading to a low (20%) rate of transcribed words. In the following presentation after a 58 second Q&A break, speakers exhibit strong accents, resulting in a larger than usual WER. The combination of these factors causes the large error in phoneme alignment of up to 80 seconds.

Our experimental setup and calculation of phoneme alignment error causes an additional error that is not measured here. Ground truth alignment values are manually inserted in the transcripts. For the manual transcription, time codes are placed every 10 seconds, and for automatic transcriptions at varying points between 10 and 30 seconds. Timestamps of words between inserted time markers are interpolated. Clearly, speech does not exhibit constant temporal intervals between words - this approximation is required because alignment cannot be clearly determined due to missing and falsely identified phrase segments. Overall the results are promising considering that large portions of text are unrecognizable, and only their phonetic constructs hint at their intended meaning.

## 5. CONCLUSION

We have presented an approach for temporally aligning highly imperfect transcripts to their original speech signal. Speech signal and transcription text are segmented into phonemes, which are then aligned with a dynamic programming edit-distance. We have demonstrated the approach on 4 datasets, showing good results, in particular for transcripts with high WER. Future investigations include extending the single global alignment to subsequent local alignments. We observe in our experiments that regions of prolonged speech not transcribed by the ASR engine cause significant errors in alignment, which are not easily recovered by the edit distance algorithm. We have already determined that the inclusion of fricatives significantly improved the alignment results; extending the phoneme detection to include additional phonemes may further improve the alignment results.